\documentclass[conference]{IEEEtran}
\IEEEoverridecommandlockouts
\usepackage{cite}
\usepackage{amsmath,amssymb,amsfonts}
\usepackage{algorithmic}
\usepackage{graphicx}
\usepackage{textcomp}
\usepackage{xcolor}
\usepackage{booktabs}
\usepackage{pgfplots}
\usepackage{lipsum} %
\usepackage{url} 

\def\BibTeX{{\rm B\kern-.05em{\sc i\kern-.025em b}\kern-.08em
    T\kern-.1667em\lower.7ex\hbox{E}\kern-.125emX}}
\begin{document}

\title{\vspace*{15pt}ASIL-Decomposition Based Resource Allocation Optimization for Automotive E/E Architectures\\
\thanks{This research was accomplished within the project AUTOtech.agil (FKZ 01IS22088x). We acknowledge the financial support for the project by the German Federal Ministry of Education and Research (BMBF).}
}

\author{\IEEEauthorblockN{Dorsa Zaheri}
\IEEEauthorblockA{\textit{Institute of Automotive Engineering (IFS)} \\
\textit{University of Stuttgart}\\
Stuttgart, Germany \\
dorsa.mohammad-zaheri@ifs.uni-stuttgart.de}
\and
\IEEEauthorblockN{Hans-Christian Reuss}
\IEEEauthorblockA{\textit{Institute of Automotive Engineering (IFS)} \\
\textit{University of Stuttgart}\\
Stuttgart, Germany \\
hans-christian.reuss@fkfs.de}
}

\maketitle

\begin{abstract}
Recent years have brought a surge of efforts in rethinking the vehicle’s electrical and/or electronic (E/E) architecture as well as the development process to reduce complexity and enable automation, connectivity, and electromobility. Resource allocation is an important step of the development process that can influence the quality of the designed system. As the design space is large and complex, intuitive design can turn into a time consuming process with sub-optimal solutions. Here, we present an approach to automatically map software components to available hardware resources. Compared to existing frameworks, our method provides a wider range of safety analyses in compliance with the ISO 26262 standard, encompassing aspects such as reliability, task scheduling, and automotive safety integrity level (ASIL) compatibility. We propose an integer linear programming (ILP)-based approach to perform ASIL decomposition technique specified by the standard. This technique proves beneficial when suitable hardware resources are unavailable for a safety-critical application. We formulate a multi-objective optimization problem to minimize both the development cost and the maximum execution times of critical function chains. The effectiveness of the proposed approach is investigated on an exemplary case study, and the results of the performance analysis are presented and discussed.
\end{abstract}


\section{Introduction}
The automotive industry is witnessing massive structural changes driven by recent technological advancements. The advent of advanced driver assistance systems (ADAS) features and autonomous driving capabilities has led to an increased number of functions and electronic systems, thereby intensifying the design complexity of automotive architectures. In response to these challenges, the concept of the software-defined vehicle (SDV) was introduced and became increasingly prevalent. Automakers are reconsidering electrical and/or electronic (E/E) architectures and adopting new methodologies for designing, analyzing, and managing automotive systems. As studies show, the industry is shifting towards zonal and central architectures \cite{ref1, ref2,ref28}. Next-generation vehicles are expected to consist of powerful general-purpose computing units and high-bandwidth in-vehicle communication protocols \cite{ref3,ref4}.

The aforementioned technological innovations are giving rise to the demand for functional safety. As a result, safety requirements must be carefully considered and addressed during the design phase of future vehicles. For this reason, the automotive industry has been gradually moving toward compliance with new standards, including ISO 26262.

ISO 26262, titled 'Road vehicles - Functional Safety', is an adaptation of the functional safety standard IEC 61508 for the automotive domain, which applies to safety-related E/E systems. This standard, defined by the international organization for standardization (ISO) in 2011 and revised in 2018, includes guidance to avoid failures or malfunctions of E/E components by establishing suitable requirements and processes \cite{ref5}. The safety life cycle outlined in ISO 26262 can be applied at the vehicle, system, hardware, and software levels. In the initial phases of development, hazards arising from faults are identified. This analysis is then used to determine the required ASIL level for each item. So far, ensuring ISO 26262 compliance has been a time-consuming process primarily done manually.

The SDV relies on the separation of software functionalities from physical hardware \cite{ref7}. As a result, a pivotal stage with a high impact on the quality of the designed system is the process of mapping software to available hardware resources, commonly referred to as resource allocation or deployment in literature \cite{ref8}. According to ISO 26262, safety requirements such as reliability, freedom from interference (FFI), ASIL compatibility and timing constraints must be met when linking software to hardware components. Therefore, the deployment process becomes more challenging due to conflicting constraints and increasing architectural complexity. Employing an optimization algorithm can effectively address this challenge and automate the task.

In this work, we present an optimization framework that automates the allocation of software components to hardware resources for automotive E/E architecture while considering safety requirements. Software architecture, which captures system functionalities, is represented using graphs. The hardware architecture may include single-core or multi-core computing units. Based on ISO 26262, a set of safety constraints is derived and embedded into a mixed-integer linear programming (MILP) model. Our proposed model performs ASIL decomposition when suitable hardware resources are not available. We formulate a multi-objective optimization problem to generate deployment solutions that minimize development costs and the execution time of critical function chains. We demonstrate results obtained by our approach including mapping, scheduling, and ASIL decomposition schemes, on an exemplary case study.

The content of this paper is organized as follows: In Section II, we review previous studies on designing and optimizing automotive E/E architectures. Section III describes our methodology and system model in more detail. Safety, as well as non-safety constraints, are presented in Section IV. The proposed multi-objective optimization problem is also discussed in this section. Section V presents the deployment results obtained by applying this method to an exemplary case study. Furthermore, we conduct a comprehensive evaluation of our approach and compare the results with those achieved using a genetic algorithm (GA), which is also discussed in this section. Finally, areas for further research are identified in Section VI.

\section{Related Work}
Our paper touches on two different research topics: resource allocation optimization and ASIL decomposition. In this section, we present the publications most relevant to our work.  

The authors of \cite{ref8} employed a domain-specific language to enable the optimization of automotive E/E architectures while adhering to various constraints. Among these constraints, they specifically addressed the ASIL compatibility between software and hardware components; however, they have left the automatic decomposition of ASIL levels for future work. Meedeniya et al. \cite{ref10} used a genetic algorithm to address the problem of reliability-driven component deployment for embedded systems. Our work includes additional safety constraints and follows a different solving strategy to ensure the fulfillment of reliability requirements. In \cite{ref11} a framework for designing and optimizing E/E architecture is presented.  In contrast to our work, where ASIL levels are considered for each application, here, applications are simply categorized as safe or non-safe, without taking reliability requirements into account. Kampmann et al. \cite{ref12} proposed an approach to automate resource allocation, with a focus on optimizing power consumption and scheduling parameters. Like the methodologies outlined in \cite{ref11} and our approach, a linear model is employed to tackle this issue. This study extends upon the service-oriented software architecture presented in \cite{ref13}. Scheduling and deployment of mixed-criticality multicore architectures considering core numbers and ASIL level of tasks is discussed in \cite{ref14}. The study demonstrated the viability of using SMT solvers for generating deployments and schedules, but did not consider ASIL decomposition. The authors in \cite{ref15} used an ILP-based approach to address resource allocation problems in the central computing platform of SDVs. The goal of the optimization problem was to find the minimum number of virtual machines of the same physical hardware under various constraints. Different to our work, the task scheduling optimization and ASIL decomposition are not considered. They extended their work by proposing a model-based approach to further automate and simplify the design process of SDVs \cite{ref16}.  

The literature mentioned above provided various methodologies aimed at solving automotive resource allocation problems. However, the integration of ASIL decomposition strategies has not been comprehensively addressed in the discussed works. ASIL decomposition can be used to reduce the criticality of safety requirements, especially in response to the increasing demand for stringent functional safety requirements within autonomous vehicle systems. There are various efforts discussing the automation of ASIL decomposition techniques during different phases of the V-model. The authors in \cite{ref17} presented two heuristic algorithms to minimize development costs and satisfy reliability goal by using ASIL decomposition during the design phase of automotive systems. This work is further extended by proposing a two-stage solution, which addresses real-time and reliability requirements simultaneously, as discussed in \cite{ref18}. Our approach enables solving a similar problem in a single-step by formulating a linear model. Hu et al. \cite{ref19} proposed a heuristic to find a low development-cost schedule with guaranteed safety. Their approach first makes use of ASIL decomposition to build a schedule with minimal development cost and then improve the reliability. The authors modeled the functionality as a directed acyclic graph and evaluated the approach on a real-life automotive benchmark. However, this work is limited to systems where ASIL decomposition is feasible. To overcome this limitation, \cite{ref20} introduced a hardware architecture exploration framework that integrates safety mechanisms into system design, enabling the achievement of the required ASIL. In \cite{ref21} and \cite{ref22}, a bottom-up and an implementation-level ASIL decomposition technique are proposed, respectively. The automotive system is described with a 3-layer model and a common-cause fault analysis is performed to validate the decomposition. Dhouibi et al. \cite{ref23} introduced a methodology for ASIL allocation and decomposition through a system of linear equations, offering the capability of generating multiple solutions. While our approach to ASIL decomposition draws inspiration from this work, we diverge slightly in methodology; whereas they use fault trees to determine ASIL allocation, our work assumes pre-existing ASIL allocations to functions and focuses solely on decomposing ASILs in order to optimize resource allocation.
To the best of our knowledge, no study has yet integrates ASIL decomposition and resource allocation into a single-step process to produce solutions that concurrently minimize development cost and latency while ensuring compliance with safety requirements. 

\section{Methodology}
In this section, we discuss our resource allocation approach and system model.  As an input to our model, we have a set of mixed-critically applications, which consists of functional blocks (also referred to as tasks) and their ASIL requirements, a set of hardware components (ECUs) and their properties, and a set of requirements that must be fulfilled. Our objective is to decompose tasks and map them to processing elements such that the overall development cost and latency are minimized and safety requirements are satisfied. To achieve this, we formulate the relevant information and constraints into an MILP model, which is then processed by the solver. If necessary, the solver applies decomposition and determines the optimal mapping decisions to ensure that safety requirements are met while minimizing overall development cost and latency. Given the considerable cost of processors with high ASIL levels, their utilization is ideally reserved for executing tasks demanding a very high level of safety. Consequently, our approach employs the ASIL decomposition technique only for tasks that have a higher ASIL level compared to all available ECUs. As part of this process, an ASIL compatibility assessment is conducted for all tasks, prior to the allocation of resources. The ASIL decomposition constraints are applied to a given task only if no ECU is capable of fulfilling the ASIL compatibility requirement for that task. Alternatively, our proposed approach also facilitates the application of ASIL decomposition to all tasks. Further details on this methodology are provided in the next section.

\subsection{System Model}
In terms of hardware, an automotive hardware architecture is considered as a distributed system consisting of heterogeneous computing units, interconnected by in-vehicle communication networks. In terms of software, an application can be represented as a directed acyclic graph (DAG), where its functional blocks can be implemented into several computing units. In the following, the hardware and software model, reliability calculation as well as ASIL decomposition scheme defined in ISO 26262 are introduced in detail. The notations used in this work are summarized in Table~\ref{tab1}.

\subsubsection{Software Model}
Mathematically, the software architecture is represented as a graph $G (T, D)$, where vertices $T$ represent tasks, which are subcomponents of applications. Edges $D$ represent dependencies between tasks. We utilize the binary parameter $d_{i,j}$ to signify the dependency between task $T_i$ and $T_j$. If task $T_i$ needs to be performed before task $T_j$, then $d_{i,j}=1$. As shown in Table~\ref{tab1}, we denote a set of tasks as $T=\{T_1, \ldots, T_n\}$. The ASIL level of task $T_i$ is shown by ${\Lambda}_{T_i}$. The set of tasks with ASIL level greater than any available ECUs is denoted as $T' \subseteq T$. ASIL decomposition constraints are exclusively applied to tasks within the set $T'$. It should be noted that setting $T' = T$ subjects all tasks to decomposition. As the tasks might be decomposed, we use ${m}_{T_{i,h}}$ for representing the memory, expressed in MB (megabyte), of the task $T_i$ with its corresponding ASIL level h. We defined the worst-case execution time (WCET) of $T_i$ running on hardware component $E_k$ with ASIL h as ${wcet}_{i,k,h}$.  

\subsubsection{Hardware Model} 
The hardware model consists of a few general-purpose ECUs. Although real architectures contain sensors and actuators, we are not defining them in our model. This is because the mapping of tasks to sensors and actuators is not meaningful. However, their effect on the deployment process is considered as localization constraint. As can be seen in Table~\ref{tab1}, we denote a set of all ECUs as $E = \{E_1, \ldots, E_m\} $. The binary parameter ${loc}_{i,k}$ is used to describe the feasibility of executing task $T_i$ on  $E_k$. If task $T_i$ can be deployed to $E_k$ then ${loc}_{i,k}=1$, otherwise ${loc}_{i,k}=0$.  The failure rate of the $E_k$ is defined as ${\lambda}_{k}$. The ASIL level that $E_k$ supports is considered as ${\Lambda}_{E_k}$. The size of its memory, expressed in MB (megabyte), is defined as ${m}_{E_k}$. To model communication network, the worst-case response time (WCRT) of a communication message sent from $T_i$ to $T_j$ is denoted as ${wcrt}_{i,j}$.

\subsection{ASIL Decomposition Model}
\begin{figure}[t!]
	\centerline{\includegraphics[width=\columnwidth]{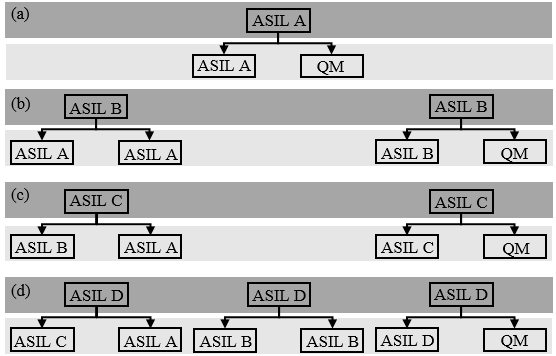}}
	\caption{ASIL decomposition schemas in ISO 26262 \cite{ref24}}
	\label{Fig1}
\end{figure}

According to ISO 26262, four ASIL levels, from ASIL A to ASIL D, are defined to represent the stringency of safety requirements. ASIL A represents the lowest criticality level, while ASIL D dictates the highest. There is another level called quality management (QM) which denotes non-safety-critical items that do not need any safety requirements. ASIL decomposition technique can be used to reduce the required ASIL level of a requirement by dividing it into multiple redundant requirements, each with a lower ASIL level \cite{ref24}. Fig.~\ref{Fig1} illustrates the ASIL decomposition schemas introduced by ISO 26262. In total, for an ASIL D requirement five decomposition options can be used:  ASIL C+ ASIL A, ASIL B + ASIL B, ASIL B + 2 × ASIL A, 4 × ASIL A and ASIL D + QM. In order to formalize the decomposition schemas defined by the standard, a numerical value is assigned to each of ASIL levels: A = 1, B = 2, C = 3, D = 4.  As discussed in \cite{ref23}, the decomposition pattern is verified if the sum of the ASIL values of the decomposed requirements is equal to the original ASIL value. 
The potential decomposition schemes are derived from (\ref{eq1}). The coefficient $\alpha_{i,h}$, where \( h \in [1\ldots 4] \), denotes the number of redundant subtasks with ASIL h for $T_i$. This is formulated in our MILP model as (\ref{eq9}). The final values of $\alpha_{i,1}$ through $\alpha_{i,4}$ indicate the decomposition scheme for each task.

\begin{equation}
\underbrace{
\begin{pmatrix}
\alpha_{11} & \cdots & \alpha_{14} \\
\vdots & \ddots & \vdots \\
\alpha_{n1} & \cdots & \alpha_{n4}
\end{pmatrix}
}_{\alpha_{i,h}}
\times
\begin{pmatrix} 
1 \\
2 \\
3 \\
4
\end{pmatrix}
=
\begin{pmatrix}
\Lambda_{T_1} \\
\vdots \\
\Lambda_{T_n}
\end{pmatrix}
\label{eq1}
\end{equation}

\subsection{Reliability Calculation}
According to ISO 26262-9 \cite{ref24}, qualitative and quantitative safety analyses should be performed during the concept and product development phases. One of the hardware architecture metrics introduced by the standard is the probabilistic metric for random hardware failures (PMHF). PMHF is a safety metric that represents a quantitative analysis evaluating the violation of the safety goal by random failures of the hardware elements. Table\,\ref{tab2} lists the target values of PMHF under different ASILs according to ISO 26262-5. To perform safety analysis, we calculate the probability of failure (PoF) and compare the quantitative result of the analysis with a target value. The probability of failure is complementary to reliability and can be calculated using (\ref{eq2})  \cite{ref25}: 

\begin{equation}
\text{PoF}(t) = 1 - R(t) = 1 - e^{-\lambda \cdot t}  
\label{eq2}
\end{equation}

In the above, $\lambda$ is the constant failure rate per time unit for an ECU and $t$ specifies the system’s life time. If ASIL decomposition is applied, as defined in ISO 26262-9, the evaluation of the hardware architectural metrics and the assessment of safety goal violations due to random hardware failures remain unchanged. This means, for example, if we decompose an ASIL D task into B(D) + B(D) and allocate these to different resources, both decomposed subtasks should fulfill the hardware metrics for the ASIL before decomposition, i.e. ASIL D. We use (\ref{eq3}) to calculate the overall probability of failure for a task $T_i \in T'$ after decomposition:
\begin{equation}
\text{PoF}_{i} = \prod_{\substack{h\in[1..4],\\ E_{k}\in E}}\left(1 - e^{-\lambda_{k} \cdot t} \cdot x_{i,k,h}\right), \quad \forall T_{i} \in T^{\prime}
\label{eq3}
\end{equation}

In this study, we assume $t=5000 \, h$ as specified by ISO 26262-2011 Part 10 Annex B.4. The variable $x_{i,k,h}$ is a binary decision variable and refers to the mapping of task $T_i$ with ASIL h on $E_k$. Thus, a task $T_i$ is deployed to ECU $E_k$ with ASIL h if $x_{i,k,h} =1$.  To validate the fulfillment of the PMHF requirement (\ref{eq4}) is used:

\begin{equation}
\text{PoF}_{i} \leq 1 - e^{-\lambda_{\text{tar}} (\Lambda_{T_i}) \cdot t}, \quad \forall T_{i} \in T^{\prime} 
\label{eq4}
\end{equation}

The $\lambda_{\text{tar}} (\Lambda_{T_i})$ is the random hardware failure target value for task $T_i$ with ${\Lambda}_{T_i}$ derived from Table\,\ref{tab2}. It is worth noting that, due to the parallel nature of decomposed subtasks, we calculate the PoF for these subtasks using parallel system formulas commonly employed in reliability calculation. Equation (\ref{eq4}) represents a nonlinear constraint and hence cannot be integrated into our MILP model. The equivalent linear equation to (\ref{eq4}) is derived by taking logarithms, yielding  (\ref{eq5}):
\begin{equation}
\begin{split}
&\forall T_{i} \in T^{\prime}:\\
&\sum_{\substack{E_{k} \in E \\ h\in[1\ldots4]}} \ln (1 - e^{-\lambda_{k} \cdot t} \cdot x_{i,k,h} ) \leq \ln (1 - e^{-\lambda_{\text{tar}} (\Lambda_{T_i}) \cdot t} ) 
\label{eq5}
\end{split}
\end{equation}
Due to the fact that $x_{i,k,h}$ is a binary variable, we can replace $\ln \left(1 - e^{-\lambda_{k} \cdot t} \cdot x_{i,k,h} \right)$ with $\ln \left(1 - e^{-\lambda_{k} \cdot t} \right) \cdot x_{i,k,h}$. This is formulated in our model as (\ref{eq24}).

\begin{table}[h!]
\renewcommand{\arraystretch}{1} 
\caption{Summary of Notations}
\centering
\begin{tabular}{p{0.45\linewidth}p{0.45\linewidth}} 
\toprule
\multicolumn{2}{c}{\textbf{Hardware components}} \\
\midrule
$E = \{E_1, \ldots, E_m\} $ & Set of all ECUs\\
${\Lambda}_{E_k}$& ASIL level of \(E_k \in E\) \\
${loc}_{i,k}$& Feasibility of running $T_i$ on $E_k$\\
${m}_{E_k}$& Memory of $E_k$\\
${\lambda}_{k}$& Failure rate of $E_k$\\
${wcrt}_{i,j}$& WCRT of a communication message sent from $T_i$ to $T_j$\\
\midrule
\multicolumn{2}{c}{\textbf{Applications}} \\
\midrule
$A$ & Set of all applications,  \(A_p \in A\)\\
$T = \{T_1, \ldots, T_n\} $& Set of all tasks of applications\\
$T'$ & Set of tasks that require ASIL decomposition $T' \subseteq T$\\
${\Lambda}_{T_i}$& ASIL level of \(T_i\in T\)\\
${d}_{i,j}$& Dependency between $T_i$ and $T_j$\\
${m}_{T_{i,h}}$& Memory of $T_i$ with ASIL $h^*$\\
${wcet}_{i,k,h}$& WCET of $T_i$ runnig on $E_k$ with ASIL h\\
${c}_{i,k,h}$& Development cost of $T_i$ on $E_k$ with ASIL h\\
\midrule
\multicolumn{2}{c}{\textbf{Decision variables}} \\
\midrule
${x}_{i,k,h}$& Mapping of $T_i$ with ASIL h on $E_k$\\
${\alpha}_{i,h}$& Number of redundant subtasks with ASIL h for $T_i$\\
${\tau}_{i,k}$& Starting time of $T_i$ on $E_k$\\
${\Phi_p}$& Maximum execution time of $A_p$\\
${\theta}$& Support variable \\
\bottomrule
\multicolumn{1}{c}{$^*h$ is an integer in the range [1\ldots 4]} \\
\end{tabular}
\label{tab1}
\end{table}

\begin{table}[ht]
\centering
\caption{Random hardware failure target values}
\label{tab2}
\begin{tabular}{cc}
\toprule
\textbf{ASIL} & \textbf{Random hardware failure target values} \\
\midrule
D & $ <10^{-8} \, \text{h}^{-1} $ \\
C & $ <10^{-7} \, \text{h}^{-1} $ \\
B & $ <10^{-7} \, \text{h}^{-1} $ \\
\bottomrule
\end{tabular}
\end{table}

\section{MILP Model Formulation}
We formulate a single-step multi-objective optimization problem to automate resource allocation and ASIL decomposition. The objective goals are set to minimize both development costs and the maximum execution time of an application while ensuring the fulfillment of safety requirements. We consider task dependencies and compute non-preemptive schedules accordingly. The resource allocation problem is formulated as follows:

\begin{equation}
\text{min} \quad \sum_{\substack{T_{i} \in T,\\ E_{k} \in E,\\ h \in [1\ldots4]}} \text{c}_{i,k,h} \cdot \text{x}_{i,k,h} 
\label{eq6}
\end{equation}

\begin{equation}
\text{min} \quad \Phi_p
\label{eq7}
\end{equation}

subject to:

\begin{equation}
\sum_{\substack{E_{k} \in E}} x_{i,k,h} = 1, \quad \forall T_{i} \in T: T_{i}\notin T^{\prime} ; h=\Lambda_{T_i}
\label{eq8}
\end{equation}

\begin{equation}
\sum_{h=1}^{4} \alpha_{i,h} \times h = \Lambda_{T_i}, \quad \forall T_{i} \in T^{\prime} 
\label{eq9}
\end{equation}

\begin{equation}
\sum_{E_{k} \in E} x_{i,k,h} = \alpha_{i,h}, \quad \forall T_{i} \in  T^{\prime}; h \in [1\ldots 4] 
\label{eq10}
\end{equation}

\begin{equation}
\sum_{h=1}^{4} x_{i,k,h} \leq 1, \quad \forall E_{k} \in E; T_{i} \in T^{\prime}
\label{eq11}
\end{equation}

Equations (\ref{eq6}) and (\ref{eq7}) are used to minimize the overall development cost and the maximum execution time of a particular application $(A_p \in A)$, respectively. As mentioned in the last section, $x_{i,k,h}$ is a decision variable and $x_{i,k,h}=1$ if task $T_i$ is deployed on $E_k$ with ASIL level h. According to the ISO 26262 standard, all software components executed on the same hardware element must comply with the highest ASIL among them unless evidence demonstrating fulfillment of coexistence criteria, specifically FFI, is provided. This implies that the hardware architecture must incorporate appropriate separation mechanisms to support mixed-criticality applications. For simplicity, this model assumes these criteria are satisfied, allowing mixed-criticality tasks to coexist. 
\\Constraint (\ref{eq8}) ensures that each task, which is not subjected to decomposition, is mapped to exactly one ECU. Possible decomposition patterns are provided using (\ref{eq9}). As an example, if task $T_i$ with ${\Lambda}_{T_i}=4 \, (ASIL\, D)$ requires decomposition, one possible solution is to replace it with two redundant ASIL B(D) subtasks. This is implied by $\alpha_{i,2} =2$ and $\alpha_{i,1}=\alpha_{i,3}=\alpha_{i,4}=0$.  Constraint (\ref{eq10}) is used to map decomposed subtasks according to the pattern derived from (\ref{eq9}). In our example, it means that the two decomposed subtasks with ASIL B should be mapped on two ECUs. In order to guarantee FFI and avoid deploying redundant subtasks on a same ECU, (\ref{eq11}) is used. 

\begin{equation}
x_{i,k,h} = 0, \quad \forall T_{i} \in T; E_{k} \in E; h \in [1\ldots 4]: \text{loc}_{i,k} = 0 
\label{eq12}
\end{equation}

The localization constraint is applied using (\ref{eq12}). It ensures that a task is mapped to an ECU only if the necessary peripheral for that task is present. Constraint (\ref{eq13}) is used to guarantee that ECUs possess sufficient memory capacity to host all assigned tasks.

\begin{equation}
\sum_{\substack{h \in [1 \ldots 4], \\ T_{i}\in T}} m_{T_{i,h}} \cdot x_{i,k,h} \leq m_{E_{k}}, \quad \forall E_{k} \in E 
\label{eq13}
\end{equation}

ASIL compatibility is checked for generated deployments using (\ref{eq14}), (\ref{eq15}), depending on whether decomposition is applied to the task or not. For clarity, the parameters $\Lambda_{T_i}$ and $\Lambda_{E_k}$ represent integer values ranging from 1 to 4, where ASIL D corresponds to 4 and ASIL A corresponds to 1.
\begin{equation}
x_{i,k,h} \cdot \Lambda_{T_i} \leq \Lambda_{E_k}, \quad \forall E_{k} \in E; T_{i} \notin T^{\prime}; h = \Lambda_{T_i}
\label{eq14}
\end{equation}

\begin{equation}
x_{i,k,h} \cdot h \leq \Lambda_{E_k}, \quad \forall E_{k} \in E; T_{i} \in T^{\prime}; h \in [1\ldots 4]
\label{eq15}
\end{equation}

Constraints (\ref{eq16}) to (\ref{eq22}) are used to schedule tasks. We used ${\tau}_{i,k}$ to denote the start time of task/subtask $T_i$ running on $E_k$.  If $T_i$ should be performed before $T_j$ (i.e., $d_{i,j}=1$), then the start time of $T_j$ must be no earlier than the finish time of $T_i$ if they are allocated to the same ECU. This is ensured using (\ref{eq16}). In case that the tasks/subtasks are allocated to different hardware elements, the WCRT must be also included as in (\ref{eq17}).
\begin{equation}
\begin{aligned}
& \forall (T_{i}, T_{j}) \in T;  E_{k} \in E:  T_{i} \neq T_{j}; d_{i,j}= 1:\\
& \tau_{i,k} +\sum_{\substack{h_1 = 1 \\ h_2 = 1}}^{4} x_{i,k,h_{1}} \cdot wcet_{i,k,h_{1}} \cdot x_{j,k,h_{2}} \leq \tau_{j,k}\\
\end{aligned}
\label{eq16}
\end{equation}

\begin{equation}
\begin{aligned}
& \forall (T_{i}, T_{j}) \in T; (E_{k}, E_{m}) \in E:  T_{i} \neq T_{j}; E_{k} \neq E_{m}; d_{i,j}= 1: \\
& \tau_{i,k} +\sum_{\substack{h_1 = 1 \\ h_2 = 1}}^{4} x_{i,k,h_{1}} \cdot (wcet_{i,k,h_{1}}+wcrt_{i,j}) \cdot x_{j,m,h_{2}} \leq \tau_{j,m}\\
\end{aligned}
\label{eq17}
\end{equation}

For tasks $T_i$ and $T_j$ running on the same ECU and $d_{i,j}=0$, following condition should be satisfied:

\begin{equation}
\begin{aligned}
\forall (T_{i}, T_{j}) \in T; &\,E_{k} \in E:  T_{i} \neq T_{j};  d_{i,j}= 0:\\
\tau_{i,k} + \sum_{h=1}^{4}&wcet_{i,k,h}  \cdot x_{i,k,h} \leq \tau_{j,k}  \\
& \text{or}& \\
 \tau_{j,k} + \sum_{h=1}^{4}  &wcet_{j,k,h} \cdot x_{j,k,h} \leq \tau_{i,k}
\label{eq18}
\end{aligned}
\end{equation}

The above constraint is an either-or condition that need to be converted to linear format. We used the Big M method by introducing a binary decision variable ${\theta_{i,j}}$. Our approach is inspired by \cite{ref26} and is encoded as (\ref{eq19})-(\ref{eq22}). The constant M is set at a large enough value.

\begin{align}
\forall (T_{i}, T_{j}) \in T; & \, E_{k} \in E:  T_{i} \neq T_{j}; d_{i,j}= 0: \notag \\
\tau_{i,k} - \tau_{j,k} - &\sum_{h=1}^{4} wcet_{j,k,h} \cdot x_{j,k,h} \geq -M \cdot \theta_{i,j} \label{eq19} \\
 \tau_{i,k} - \tau_{j,k} - &\sum_{h=1}^{4} wcet_{j,k,h} \cdot x_{j,k,h} \leq M \cdot (1 - \theta_{i,j}) \label{eq20}
\end{align}

\begin{equation}
\begin{split}
\forall (T_{i}, T_{j}) \in T;  &\, E_{k} \in E; \, h_{1}, h_{2} \in [1\ldots 4]: \\
& x_{i,k,h_{1}} + x_{j,k,h_{2}} + \theta_{i,j} + \theta_{j,i} \leq 3 
\label{eq21} 
\end{split}
\end{equation}

\begin{equation}
\theta_{i,j} = 1, \quad \forall (T_{i}, T_{j}) \in T: \, d_{i,j}= 1
 \label{eq22}
\end{equation}

The maximum execution time of an application $A_p \in A$ is calculated using (\ref{eq23}). We aim to minimize this value as formulated in (\ref{eq7}).

\begin{equation}
\begin{aligned}
\forall T_{i} \in (T \cap A_{p}); & E_{k} \in E; A_{p} \in A:\\
\quad & \Phi_{p} \geq \tau_{i,k} + \sum_{h=1}^{4} wcet_{i,k,h} \cdot x_{i,k,h}  
\end{aligned}
\label{eq23}
\end{equation}

Finally, Constraint (\ref{eq24}) ensures the fulfillment of PMHF requirements for all decomposed tasks. 

\begin{equation}
\begin{aligned}
&\forall T_{i} \in T'; t=5000:\\
&\sum_{\substack{E_{k} \in E, \\ h \in [1\ldots4]}} \ln (1 - e^{-\lambda_{k} \cdot t} ) \cdot x_{i,k,h}  \leq \ln (1 - e^{-\lambda_{\text{tar}} (\Lambda_{T_i}) \cdot t} ) \\
\end{aligned}
\label{eq24}
\end{equation}

 The constraints specified in this section can be applied to any solver, such as CPLEX\footnote{\url{https://www.ibm.com}}, Gurobi\footnote{\url{https://www.gurobi.com}} and others, which are designed to efficiently handle linear constraints and objectives.

\section{Experimental Results}

\begin{figure}[htbp]
\centering
	\includegraphics[width=0.4\columnwidth]{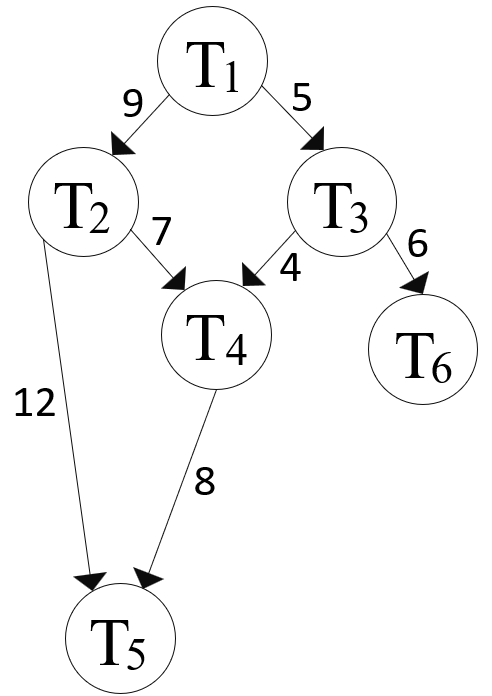}
	\caption{The exemplary graph consisting of six tasks}
	\label{Fig2}
\end{figure}

The focus of this section is to provide evaluation results for the proposed approach. To this end, we tested our formulation using an exemplary set of ECUs and an application graph depicted in Fig.\,\ref{Fig2}. The application consists of six tasks, all of which are considered to be safety-critical at ASIL D. The dependencies between tasks are represented using arrows. Weights of the edges between tasks represents the WCRT. The hardware architecture compromises four general-purpose ECUs. We assumed that all ECUs possess required peripherals for running a task; therefore $loc_{i,k}=1$ for all tasks. The properties of all ECUs are given in Table\,\ref{tab3}. The assumed WCETs of each task on different ECUs and ASIL levels are listed in Table\,\ref{tab4}. We assumed that for a given task, the WCET is higher when it has a higher ASIL compared to when it has a lower ASIL. Table\,\ref{tab5} contains the development costs of each task on different ECUs with different ASIL levels. The development cost of a task with ASIL D is considered to be greater than two decomposed subtasks with ASIL B. The required memory of the tasks is provided in Table\,\ref{tab6}.

\begin{table}[htbp]
    \caption{Properties of exemplary ECUs}
    \centering
    \begin{tabular}{ccccc}
        \toprule
        {Property} & {ECU1} & {ECU2} & {ECU3} & {ECU4} \\
        \midrule
        ${\Lambda}_{E_k}$ & C & B & B & C \\
        ${m}_{E_k}$ & 8 GB & 8 GB & 2 GB & 16 GB \\
        ${\lambda}_{k} [\, \text{h}^{-1}]$ & $10 \times 10^{-7} $ & $20  \times 10^{-7}$ & $30  \times 10^{-7} $ & $8 \times 10^{-7}$ \\
        \bottomrule
    \end{tabular}
    \label{tab3}
\end{table}

\begin{table*}
    \caption{WCETs (unit: ms) of tasks on different ECUs and ASILs}
    \centering
    \setlength{\tabcolsep}{4.3pt} 
    \begin{tabular}{ccccccccccccccccccccccccc}
        \toprule
        & \multicolumn{4}{c}{$T_1$} & \multicolumn{4}{c}{$T_2$} & \multicolumn{4}{c}{$T_3$} & \multicolumn{4}{c}{$T_4$} & \multicolumn{4}{c}{$T_5$} & \multicolumn{4}{c}{$T_6$} \\
        \cmidrule(lr){2-5} \cmidrule(lr){6-9} \cmidrule(lr){10-13} \cmidrule(lr){14-17} \cmidrule(lr){18-21} \cmidrule(lr){22-25} 
           & $E_1$ & $E_2$ & $E_3$ & $E_4$ & $E_1$ & $E_2$ & $E_3$& $E_4$ & $E_1$ & $E_2$ & $E_3$ & $E_4$ & $E_1$ & $E_2$ & $E_3$ & $E_4$ & $E_1$ & $E_2$ & $E_3$ & $E_4$ & $E_1$ & $E_2$ & $E_3$ & $E_4$ \\
        \midrule
        ASIL A & 7 & 4 & 5 & 8 & 10 & 6 & 4 & 7 & 9 & 8 & 6 & 13 & 14 & 11 & 10 & 10 & 14 & 12 & 8 & 16 & 8 & 1 & 3 & 8 \\
        ASIL B & 9 & 6 & 7 & 10 & 12 & 8 & 6 & 9 & 11 & 10 & 8 & 15 & 16 & 13 & 12 & 12 & 16 & 14 & 10 & 18 & 10 & 3 & 5 & 10 \\
        ASIL C & 11 & 8 & 9 & 12 & 14 & 10 & 8 & 11 & 13 & 12 & 10 & 17 & 18 & 15 & 14 & 14 & 18 & 16 & 12 & 20 & 12 & 5 & 7 & 12 \\
       ASIL D & 13 & 10 & 11 & 14 & 16 & 12 & 10 & 13 & 15 & 14 & 12 & 19 & 20 & 17 & 16 & 16 & 20 & 18 & 14 & 22 & 14 & 7 & 9 & 14 \\
        \bottomrule
    \end{tabular}
    \label{tab4}
\end{table*}

\begin{table*}
    \caption{Development costs of tasks on different ECUs and ASILs}
    \centering
    \setlength{\tabcolsep}{4.3pt} 
    \begin{tabular}{ccccccccccccccccccccccccc}
        \toprule
        & \multicolumn{4}{c}{$T_1$} & \multicolumn{4}{c}{$T_2$} & \multicolumn{4}{c}{$T_3$} & \multicolumn{4}{c}{$T_4$} & \multicolumn{4}{c}{$T_5$} & \multicolumn{4}{c}{$T_6$} \\
        \cmidrule(lr){2-5} \cmidrule(lr){6-9} \cmidrule(lr){10-13} \cmidrule(lr){14-17} \cmidrule(lr){18-21} \cmidrule(lr){22-25} 
           & $E_1$ & $E_2$ & $E_3$ & $E_4$ & $E_1$ & $E_2$ & $E_3$& $E_4$ & $E_1$ & $E_2$ & $E_3$ & $E_4$ & $E_1$ & $E_2$ & $E_3$ & $E_4$ & $E_1$ & $E_2$ & $E_3$ & $E_4$ & $E_1$ & $E_2$ & $E_3$ & $E_4$ \\
        \midrule
        ASIL A & 8 & 6 & 5 & 7 & 9 & 7 & 7 & 8 & 5 & 4 & 3 & 5 & 4 & 3 & 2 & 4 & 6 & 3 & 3 & 5 & 8 & 5 & 5 & 6 \\
        ASIL B & 11 & 8& 7 & 9 & 13 & 9 & 8 & 12 & 8 & 6 & 6 & 8 & 7 & 6 & 5 & 7 & 9 & 5 & 4 & 8 & 13 & 9 & 8 & 10 \\
        ASIL C & 18 & 15 & 12 & 16 & 20 & 17 & 17 & 19 & 14 & 11 & 11 & 13 & 12 & 10 & 8 & 11 & 15 & 9 & 8 & 14 & 22 & 13 & 13 & 18 \\
       ASIL D & 23 & 17 & 16 & 21 & 26 & 22 & 20 & 25 & 16 & 14 & 14 & 17 & 15 & 15 & 13 & 15 & 19 & 13 & 12 & 18 & 28 & 22 & 22 & 22 \\
        \bottomrule
    \end{tabular}
    \label{tab5}
\end{table*}

\begin{table}[htbp]
    \caption{Required memory for tasks at different ASIL levels}
    \centering
    \setlength{\tabcolsep}{5pt} 
    \begin{tabular}{ccccccc}
        \toprule
        {ASIL }& {$T_1$} & {$T_2$}& {$T_3$}& {$T_4$} & {$T_5$} & {$T_6$} \\
        \midrule
        $A$ & 2 GB & 2.5 GB & 1 GB  & 500 MB & 1 GB & 1.5 GB \\
        $B$ & 2 GB & 3 GB & 1 GB  & 1 GB & 1.5 GB & 2.5 GB \\
        $C$ & 2 GB & 3 GB & 2 GB  & 1.5 GB & 2 GB & 3 GB \\
        $D$ & 5 GB & 4 GB & 5 GB  & 3 GB & 3 GB & 4 GB \\
        \bottomrule
    \end{tabular}
    \label{tab6}
\end{table}

\begin{figure}[htbp]
\centering
	\includegraphics[width=\columnwidth]{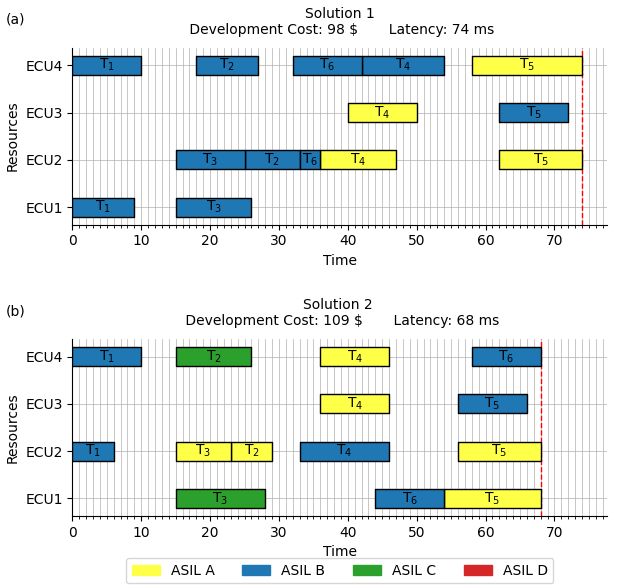}
	\caption{Optimal deployment solutions for the case study with the primary objective: (a) minimization of the development cost and (b) minimization of the maximum execution time}
	\label{Fig3}
\end{figure}

The formulated MILP problem is solved using the Gurobi solver to obtain the optimal solution. The experiments were conducted on a PC equipped with an Intel i7-8565U CPU running at 1.8 GHz, with 32 GB RAM. The optimal allocation result of the exemplary model is illustrated in Fig.\,\ref{Fig3}. In Fig. 3(a), the decomposition scheme and mapping solution are depicted, with a focus on prioritizing development costs as the primary objective. In this instance, the overall development cost amounts to 98, and the maximum execution time of the application is 74 ms. Conversely, Figure 3(b) demonstrates the allocation scenario where latency is considered as the higher-priority objective. In this case, the development cost is 109, and latency is 68 ms.

We compared the performance of the Gurobi and CPLEX solvers, widely acknowledged as the most commonly used linear solvers, in terms of computation time for our problem formulation. The solvers were executed with their default configurations. The computation time of Gurobi for solving the problem with development cost as the higher-priority objective was $0.266 \, s$, whereas CPLEX took $0.844 \, s$. In the case of prioritizing latency as the higher-priority objective, Gurobi was able to find a solution within $7.719 \, s$, while CPLEX required $58.188 \, s$. Our findings reveal an increase in computation time across all solvers as the number of tasks escalates. Based on our experiments, on average, Gurobi outperforms CPLEX in terms of solving our specific problem formulation faster. 

To further evaluate the applicability and scalability of the proposed method, we conduct a series of experiments, with cost minimization given the highest priority. Specifically, we consider a fixed hardware architecture consisting of four ECUs and analyze the solving time with respect to an increasing number of tasks deployed to these ECUs. The task graph is generated randomly using the Erdős–Rényi graph model~\cite{ref27}, with a connectivity probability of 0.9. The ECU failure rates are similar to those in Table\,\ref{tab3}. The cost values are assigned randomly while ensuring that the cost hierarchy follows the constraint: $ \text{ASIL } D>C>B>A $, with the cost of ASIL D being greater than twice that of ASIL B. Additionally, WCET and WCRT are randomly selected, with WCET values ranging between 1 ms and 20 ms, and WCRT values between 1 ms and 15 ms. Fig. 4(a) demonstrates the results of applying the proposed approach to a case where no ASIL decomposition is required. In this scenario, the ASIL of all tasks is lower than the ASIL level supported by the ECUs. To investigate the impact of ASIL decomposition on solving time, we conducted an experiment, the results of which are shown in Fig. 4(b). In the first experiment, we assumed that all tasks are ASIL D and the ECUs support up to ASIL C. This scenario requires more computation time, as all ASIL decomposition schemes need to be tested by the algorithm. In the second experiment, we considered all tasks to have ASIL C and the ECUs support up to ASIL B. As shown in Fig. 4(b), the computation time for this scenario is significantly lower than in the previous case, as expected. 
In addition to the proposed ILP approach, a penalty-based GA is developed for ASIL decomposition-based resource allocation optimization, incorporating reliability constraints but not scheduling parameters. The approach uses variable-length encoding, where each gene is assigned a random decomposition scheme from a predefined library matching its original ASIL level. The fitness function minimizes the total cost, with penalties for violating reliability constraints. Initial solutions are generated randomly, and tournament selection is used to choose parents. Crossover occurs at a randomly selected point between 1 and the length of the parent, exchanging genetic material between parents. Mutation replaces a randomly chosen gene with a random compatible ASIL scheme. Results of applying the proposed ILP method and GA on randomly generated graphs are shown in Fig. 4(c). As expected, ILP finds the exact minimum, whereas GA does not. However, a precise timing comparison is not feasible since ILP runs only once, while GA's computation time varies significantly with the number of generations.

\begin{figure}[htbp]
\centering
	\includegraphics[width=\columnwidth]{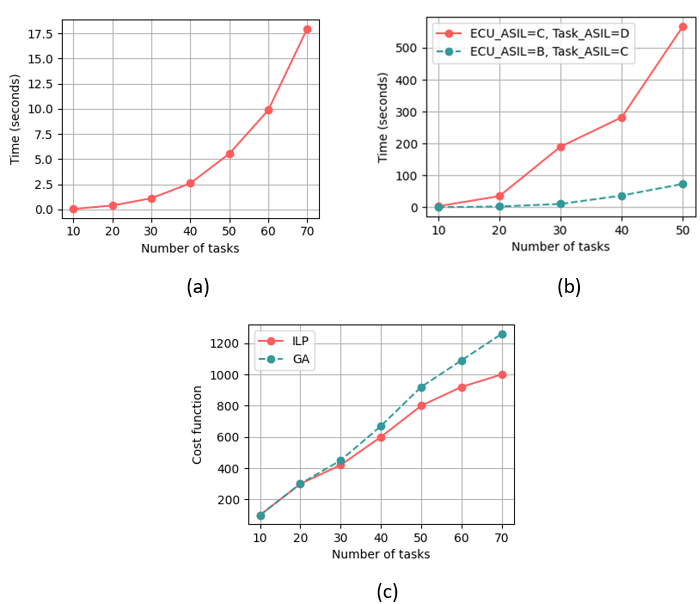}
	\caption{Performance evaluation of the proposed approach. (a) Application on a topology consisting of four ECUs with ASIL C and ASIL B tasks, without ASIL decomposition. (b) Impact of ASIL decomposition on solving time. (c) Comparison of cost function results between the proposed ILP and GA methods.}
	\label{Fig4}
\end{figure}

\section{Conclusion}
This paper proposes an approach to optimize resource allocation for automotive E/E architecture by using ASIL decomposition technique specified in ISO 26262 standard. We formulate safety requirements, such as reliability and FFI, as well as deployment problems, as an mixed-integer linear program. The proposed MILP model aims to perform ASIL decomposition and optimize the mapping of tasks and replicas to the available hardware components in a single-step. We have set up a multi-objective mixed-integer optimization model to minimize both the overall development cost and the maximum execution time of an application. The model also allows for the assignment of priority to objectives, facilitating the examination of the trade-off between them. The presented experiments and exemplary case study demonstrate the efficiency of our proposed framework. Looking ahead, we aim to expand upon our work by integrating a message routing and scheduling model into the existing MILP framework.

\vspace{12pt}

\end{document}